\documentclass[12pt, peerreviewca, onecolumn]{IEEEtran}

\usepackage{epsfig}
\usepackage{amsmath}
\usepackage{graphics}
\usepackage{amsfonts}
\usepackage{amssymb}
\usepackage{subfigure}
\usepackage{booktabs}
\usepackage{cite}

\newcommand{\vect}[1]{\boldsymbol{#1}}          
\newcommand{\mat}[1]{\boldsymbol{#1}}           

\newcommand{\define}{\stackrel{\Delta}{=}}  

\newcommand{\SNR}{\text{SNR}}             

\newcommand{\Pout}{P_{\rm out}}               
\newcommand{\dpeak}{d_{\rm peak}}               
\newcommand{\dout}{d_{\rm out}}               
\newcommand{\de}{d_{\rm e}}               
\newcommand{\dsb}{d_{\rm sb}(R)}               
\newcommand{\dsbrot}{d_{\rm sb}^{\rm rot}(R)}               

\newcommand{\dotle}{\:\dot\le\:}            

\newcommand{\Xc}{\mathcal{X}}

\newcommand{\CC}{\mathbb{C}}

\newcommand{\EE}{\mathbb{E}}

\newcommand{\beq}{\begin{equation}}
\newcommand{\eeq}{\end{equation}}

\DeclareMathOperator{\expect}{E}

\newcommand{\transp}[1]{{#1}^\text T}     

\newtheorem{theorem}{Theorem}

\renewcommand{\baselinestretch}{1.75}
\begin{document}
\title{Coded Modulation with Mismatched CSIT \\over Block-Fading Channels }
\author{T\`ung T. Kim and Albert Guill\'en i F\`abregas
\thanks{Manuscript received January 1, 2009; revised May 28, 2009. The material in this paper will be presented in part to the
IEEE International Symposium on
Information Theory, Seoul, Korea, June-July 2009.}\thanks{T. T. Kim is
with the Department of Electrical Engineering, Princeton University, Princeton, NJ 08544 (e-mail:
thanhkim@princeton.edu).} \thanks{A. Guill\'en i F\`abregas is with
the Department of Engineering, University of Cambridge, Cambridge
CB2 1PZ, UK (e-mail: guillen@ieee.org).}
 } \maketitle

\begin{abstract}
Reliable communication over delay-constrained block-fading channels with discrete inputs and mismatched (imperfect) channel state information at the transmitter (CSIT) is studied. The CSIT mismatch is modeled as Gaussian random variables, whose variances decay as a power of the signal-to-noise ratio (SNR). A special focus is placed on the large-SNR decay of the outage probability when power control with long-term power constraints is used. Without explicitly characterizing the corresponding power allocation algorithms, we derive the outage exponent as a function of the system parameters, including the CSIT noise variance exponent and the exponent of the peak power constraint. It is shown that CSIT, even if noisy, is always beneficial and leads to important gains in terms of exponents. It is also shown that when multidimensional rotations or precoders are used at the transmitter, further exponent gains can be attained, but at the expense of larger decoding complexity.
\end{abstract}


\section{Introduction}

Temporal power control across fading states can lead to dramatic improvement in the outage performance of block-fading channels \cite{Caire:PowerControl}. The intuition behind this phenomenon is that power saved in particularly bad channel conditions can be used in better channel realizations.
Power control over block-fading channels was originally studied under the idealistic assumptions of perfect channel state information (CSI) at the transmitter (CSIT) and Gaussian signal constellations \cite{Caire:PowerControl}. Acquiring perfect CSIT is however a challenging task due to the temporal variation of wireless media, as well as due to the processing and transmission delay. This motivates a large body of works studying fading channels under less optimistic assumptions about the CSIT; see for example \cite{Love:MIMOTutorials,Vu:PrecodingTutorials} and references therein.

This work considers a block-fading channel with \emph{discrete
input}, where the transmitter has access to a noisy version of the
CSI. Similarly to \cite{Lim:TradeoffWC}, we model the CSIT noise
as Gaussian random variables whose variances decay as a negative
power of the signal-to-noise ratio (SNR). Such a noise-corrupted
CSIT model is well motivated and studied in the literature; see
for example
\cite{Visotsky:Precoding,Jongren:BF_OSTBC,Zhou:EigenBF}. The rate
of decaying of the CSIT noise can also be related to practical
parameters in wireless systems \cite{Kim:IT_DMT_TDD}. Unlike the
constant-power variable-rate scenarios, studied e.g. in
\cite{Kamath:Globecom01,Lim:TradeoffWC}, we consider a
power-controlled constant-rate system. In sharp contrast to the
assumption of using Gaussian codebooks
\cite{Lim:TradeoffWC,Sharma:PowerAllocation,Steger:Training,Aggrawal:Asilomar08,Kim:TradeoffIT,Kim:IT_DMT_TDD},
the current work assumes that the input symbols are taken from a
\emph{discrete} distribution such as M-QAM or PSK.

Focusing on the high signal-to-noise ratio (SNR) regime, we establish the diversity gain of block-fading channels under the noisy CSIT model of interest.
Note that unlike in the diversity--multiplexing tradeoff analysis \cite{Zheng:Tradeoff} where the code rate grows with the SNR, herein we keep the constellation size to be $2^M$ at all values of the SNR and we do not let the code rate scale with the SNR.
We show that the diversity gain of coded-modulation systems can only match that provided by the ideal Gaussian codebooks
when the ratio between
the code rate and the constellation size is sufficiently small.
The results shed some light into the
interplay in the high-SNR regime between the number of receive antennas, the number of fading blocks,
the constellation size, the code rate, as well as the SNR exponent of the  CSIT noise variance  and the peak exponent constraint.

This paper is organized as follows. The system model is given in Section~\ref{section:model}. Section~\ref{section:prelim} introduces the fundamental concepts underlying our analysis. Section~\ref{section:asympt} presents our main results for the outage exponent with imperfect CSIT. Section~\ref{section:conclusions} draws our final considerations. The proofs of our results can be found in the appendices.

\section{System Model}
\label{section:model}

Consider transmission over a block-fading channel with $B$ sub-channels, where each
sub-channel has a single transmit and $m$ receive antennas. The
mutually independent channel gains $\vect h_1, \ldots, \vect h_B$
have independent and identically distributed (i.i.d.) complex Gaussian components with zero means and unit
variances.
The channel gains are constant during one fading block
but change from one block to the other according to some ergodic
and stationary Gaussian process. This models a typical delay-limited scenario in wireless communications, where
the delay constraint dictated by higher-layer applications prevents the system from fully exploiting time diversity \cite{Caire:PowerControl}.

The corresponding discrete-time complex baseband input-output relation for the $i$th sub-channel can be written as
\begin{equation}
\mat Y_i =  \vect h_i \, \sqrt{P_i} \,\transp{\vect x}_i + \mat W_i
\label{eq:model}
\end{equation}
where $\mat Y_i \in \CC^{m\times L}$ is the received signal matrix corresponding to block $i$, $\vect x_i \in \CC^L$ is the transmitted vector in block $i$,
$\transp{\vect x}$ denotes the transpose of $\vect x$,
and $\mat W_i\in \CC^{m\times L}$ denotes the complex additive white Gaussian noise whose entries are i.i.d. with zero means and unit variances. We denote the block length by $L$ and the power in block $i$ by $P_i$. Hence, a codeword corresponds to $BL$ channel uses.

We assume perfect CSI at the receiver (CSIR), i.e., the receiver
has perfect knowledge about all the channel gains and the powers
$P_i$. Furthermore, we assume that the transmitter has access to a
noisy version $\widehat{\vect h}_i$ of the true channel
realization $\vect h_i$, so that
\begin{equation}
\vect h_i = \widehat{\vect h}_i + \vect e_i,\qquad i =1,\ldots, B
\label{eq:mismatched_csit}
\end{equation}
where $\vect e_i \in \CC^m$ is the CSIT noise vector, independent of $\widehat{\vect
h}_i$, with i.i.d. Gaussian components with zero mean and variance
$\sigma^2_\text e$. This model of the CSIT has been well motivated in many
different contexts, such as in scenarios with delayed feedback,
noisy feedback, or in systems exploiting channel reciprocity
\cite{Visotsky:Precoding,Jongren:BF_OSTBC}. We further assume, as in \cite{Lim:TradeoffWC}, that
the CSIT noise variance decays as a power of the SNR
\begin{equation}
\sigma_\text e^2 = \SNR^{-\de}
\end{equation}
for some $\de>0$. Thus we consider a family of
channels where the second-order statistic of the CSIT noise varies
with SNR.  If the CSIT for example is estimated from the reverse link due to reciprocity, its quality will depend on the
SNR of reverse link and not the forward link. However, while the SNRs of the forward and reverse links are different, this difference will be fully captured by changing
the values of $\de$.
For convenience, we introduce the normalized channel gains
\begin{equation}
\bar{\vect h}_i = \frac{\sqrt 2}{\sigma_\text e}\vect h_i.
\end{equation}
Given $\widehat{\vect h}_i$ then $\bar{\vect h}_i$ is complex
Gaussian with mean $\frac{\sqrt 2}{\sigma_\text e}\widehat{\vect h}_i$
and a scaled identity covariance matrix.

Let
$\gamma_i\define\|\vect h_i\|^2$ be the fading magnitude of block $i$ and $\boldsymbol\gamma = [\gamma_1 \cdots
\gamma_B]$.  Further denote $\bar\gamma_i  \define \|\bar{\vect h}_i\|^2$,
$\hat\gamma_i
\define \|\widehat{\vect h}_i\|^2$, $\bar{\boldsymbol\gamma} \define [\bar\gamma_1
\cdots \bar\gamma_B]$ and ${\widehat{\boldsymbol\gamma}} \define
[\hat\gamma_1 \cdots \hat\gamma_B]$.

\begin{figure}[t]
\begin{center}
\includegraphics[width=0.7\columnwidth]{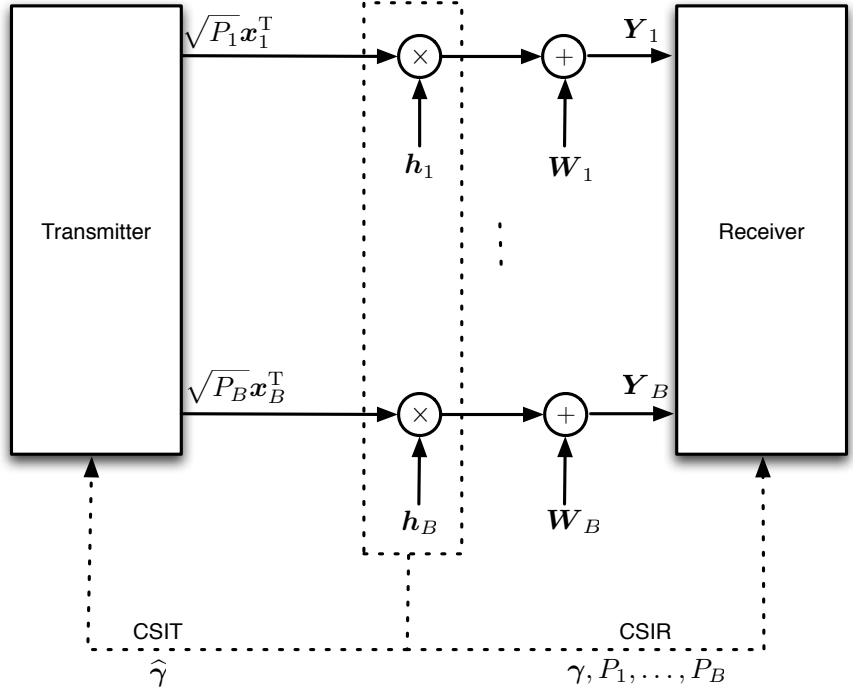}
\caption{System model and CSI assumptions.}
\label{fig:model}
\end{center}
\end{figure}

The system model and CSI assumptions are summarized in Fig. \ref{fig:model}.

\section{Preliminaries}
\label{section:prelim}

We assume transmission at a fixed-rate $R$ using a coded modulation scheme $\mathcal M \subset \CC^{BL}$ of length $BL$ constructed over a signal constellation $\mathcal X\subset \CC$ of size $2^M$ such as $2^M$-PSK or QAM. We denote the codewords of $\mathcal M$ by $\vect x = \transp{(\transp{\vect x}_1,\dotsc,\transp{\vect x}_B)}\in \CC^{BL}$. We assume that the signal constellation $\mathcal X$ has zero mean and is normalized in energy, i.e., $\EE[X]=0$ and $\EE[|X|^2]=1$, where $X$ denotes the corresponding random variable. We denote the input distribution as $Q(x)$. With these assumptions, the instantaneous input-output mutual information of the channel is given by
\begin{equation}
I(\vect \gamma) = \frac{1}{B}\sum_{i=1}^B I_{\Xc} (P_i \gamma_i)
\end{equation}
where
\begin{equation}
I_{\Xc} (s) = \EE\left[\log_2\frac{e^{-|Y-\sqrt{s}X|^2}}{\sum_{x'\in\Xc}Q(x')e^{-|Y-\sqrt{s}x'|^2}}\right]
\end{equation}
is the input-output mutual information of an additive white
Gaussian noise (AWGN) channel with SNR $s$ using uniformly the
signal constellation $\Xc$.

The outage probability is commonly defined as in \cite{Ozarow:Cellular,Biglieri:Fading}
\begin{equation}
\Pout(R) \define \Pr\{I(\vect \gamma)<R\}.
\end{equation}
In this work, we are interested in the SNR exponents of the outage probability \cite{Zheng:Tradeoff,Fabregas:CodedMod}, i.e.,
\begin{equation}
\dout \define \lim_{\SNR\to\infty} -\frac{\log \Pout(R)}{\log\SNR}.
\end{equation}
We adopt the notation $g(\SNR)\doteq\SNR^{a} \Leftrightarrow \lim_{\SNR\to\infty}\frac{\log g(\SNR)}{\log\SNR} = a$.

It has been shown in \cite{Fabregas:CodedMod,Nguyen:TightLB} that the outage exponent without CSIT is given by
\beq
\dout = m \dsb
\eeq
where
\beq
\dsb \define 1 + \left\lfloor B\left(1-\frac{R}{M}\right) \right\rfloor  = B -\left\lceil\frac{BR}{M}\right\rceil +1,
\eeq
with $\lfloor x \rfloor$ being the largest integer that is not larger than $x$ and $\lceil x \rceil$ being the smallest integer that is not smaller than $x$,
is the Singleton bound on the block-diversity of the coded modulation scheme $\cal M$ \cite{Knopp:CodedModulation,Malkamaki:CodedDiversity,Fabregas:CodedMod}. 

Due to the availability of a noisy version of the channel ${\widehat{\boldsymbol\gamma}}$, the transmitter can adapt the transmitted powers $P_i$ to the channel conditions. In this work, we consider power allocation algorithms that treat the noisy CSIT ${\widehat{\boldsymbol\gamma}}$ as if it were perfect. We consider an average power constraint, such that
\begin{equation}
\EE\left[\frac{1}{B}\sum_{i=1}^B P_i(\widehat{\boldsymbol\gamma}) \right] = \EE\left[P(\widehat{\boldsymbol\gamma})\right] \leq \SNR
\label{eq:average_pc}
\end{equation}
where we have denoted $P(\widehat{\boldsymbol\gamma}) = \frac{1}{B}\sum_{i=1}^B P_i(\widehat{\boldsymbol\gamma})$ as the instantaneous average (or normalized total) power allocated
given the noisy channel observation $\widehat{\boldsymbol\gamma}$.
Thus the SNR herein has the meaning of the average transmit power
over infinitely many fading blocks. It is well known that power allocation with average power constraints yields significant gains with respect to power allocation with peak power constraints both in terms of exponents and absolute outage probability \cite{Caire:PowerControl}. In order to give a more accurate characterization of the system behavior under practical peak-to-average power limitations, we also introduce a peak-to-average power constraint of the form
\begin{equation}
P(\widehat{\boldsymbol\gamma}) \leq \SNR^{\dpeak}
\label{eq:peak_pc}
\end{equation}
where $\dpeak$ is interpreted as the peak-to-average power SNR exponent. The case $\dpeak=1$ represents a system whose allocated power is dominated by the peak-power constraint. Asymptotically, this yields the same exponent of a system with no power control.
 By allowing $\dpeak$ to take an arbitrary value, we can model a
family of systems with different behavior in the peak power
constraint. Note that in the high-SNR regime of interest, we can for
example scale the right hand side of \eqref{eq:peak_pc} by a
constant without changing any conclusion. That is, any constant,
finite ratios between the peak and the average power provides the
same asymptotic behavior as $\dpeak=1$.

The corresponding minimum-outage power allocation rule is the solution to the following problem
\begin{equation}
\begin{cases}
\text{Minimize} &\Pout(R)\\
\text{subject to} & \EE\left[\frac{1}{B}\sum_{i=1}^B P_i(\widehat{\boldsymbol\gamma}) \right]\leq \SNR\\
& \frac{1}{B}\sum_{i=1}^B P_i(\widehat{\boldsymbol\gamma}) \leq \SNR^{\dpeak}\\
& P_i(\widehat{\boldsymbol\gamma})\geq 0, ~~i=1,\dotsc,B.
\end{cases}
\end{equation}
Solving this problem even numerically is difficult in general, given our noisy CSIT model and the discreteness of $\mathcal X$.
To date, only in the case of perfect CSIT, the minimum outage power control rule is known \cite{Nguyen:ISIT2008}, along with its asymptotic behavior. The algorithm in \cite{Nguyen:ISIT2008} would actually be used in our case by a transmitter that is ignorant of the imperfectness of the CSIT.
Nevertheless, we can characterize the
asymptotic behavior of the optimal solution in the high SNR regime.
Following the footsteps of \cite{Zheng:Tradeoff}, we note that the outage exponent of the optimal algorithm is the same as that of a power control system that allocates power uniformly across the blocks, i.e, $P_i(\widehat{\boldsymbol\gamma})=P(\widehat{\boldsymbol\gamma})$, $\forall i=1,\ldots, B$. This is because we can lower- and upper-bound the instantaneous input-output mutual information as
\begin{equation}
\frac{1}{B}\sum_{i=1}^BI_\mathcal X(P(\widehat{\boldsymbol\gamma})\gamma_i) \le \frac{1}{B}\sum_{i=1}^B I_\mathcal X(P_i(\widehat{\boldsymbol\gamma})\gamma_i) \le \sum_{i=1}^B\frac{1}{B}I_\mathcal X(BP(\widehat{\boldsymbol\gamma})\gamma_i).
\end{equation}
Since $B$ is a finite constant independent of the SNR, it does not change any asymptotic behavior of our interest.

\section{Asymptotic Behavior of the Outage Probability}
\label{section:asympt}

\subsection{Main Results}

In this section, we study the asymptotic behavior of the outage probability. In particular, our main results in terms of outage SNR exponents are stated as follows.

\begin{theorem}
Consider transmission at rate $R$ over a block-fading channel described by \eqref{eq:model} with Rayleigh fading with mismatched CSIT modeled by \eqref{eq:mismatched_csit} with inputs drawn from $\Xc$. The transmitter uses power control with an average power constraint \eqref{eq:average_pc} and a peak-to-average power constraint \eqref{eq:peak_pc}.
Then, the outage exponents are given by
\begin{equation}
d(R, \de, \dpeak) =
\begin{cases}
m  \dsb \dpeak& \dpeak \leq 1 + m\dsb\de,\\
m\dsb\left(1+m\dsb \de\right) &\dpeak > 1 + m\dsb\de.
\end{cases}
\end{equation}
\label{theorem:di}
\end{theorem}

\begin{IEEEproof}
See Appendix~\ref{apdx:main_proof}.
\end{IEEEproof}

\begin{figure}[t]
\begin{center}
\includegraphics[width=1\columnwidth]{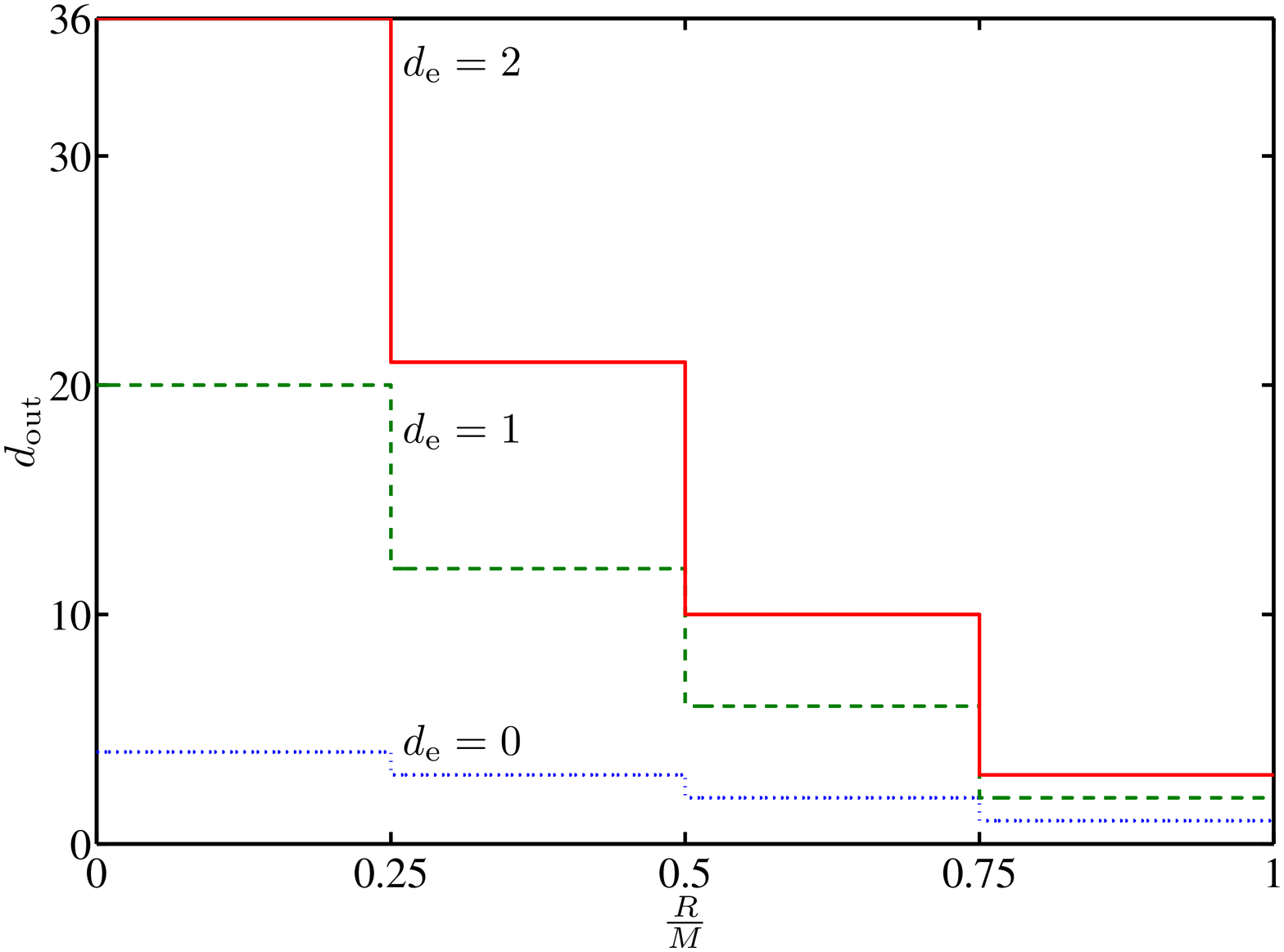}
\caption{Outage exponents for $B=4$, $m=1$ and $\dpeak > 1 + m \de\dsb$.}
\label{fig:exponents_dpeak_inf}
\end{center}
\end{figure}

To illustrate the above theorem, in Fig.~\ref{fig:exponents_dpeak_inf} we plot the outage exponents for $B=4$, $m=1$ with no CSIT (or $\de=0$) and with noisy CSIT with $\de=1,2$ when $\dpeak > 1 + m \de\dsb$. As we observe from the figure, increasing $\de$ yields a better exponent. Note that in this case, when the CSIT is perfect the exponent is infinitely large \cite{Nguyen:ISIT2008}. Observe, however, that even in the presence of imperfect CSIT, large gains are possible by using power control, with respect to the uniform power allocation case. In many practical systems we typically have  $\de<1$ and that in such scenarios $\de$
can be related to the Doppler shift\cite{Kim:IT_DMT_TDD}. In principle, achieving $\de>1$ may also be possible by means of power control in
the feedback link \cite{Steger:Training}.
Note that our main result in Theorem~\ref{theorem:di} (and Theorem~\ref{theorem:di_rot}) also holds
for nonzero-mean $\vect h_i$'s (Rician fading), because the asymptotic diversity gain only captures the
slope of the outage probability, which is the same for zero and nonzero-mean $\vect h_i$'s.

To get some insight into the problem, let us take a closer look at the results of Theorem~\ref{theorem:di} in some special cases. In the extreme case $\dpeak=1$, which implies that the average and peak power have the same exponent, we obtain $d(R, \de, 1) = m\dsb$, which is
the outage exponent for a system with short-term power control \cite{Nguyen:ISIT2008}, or no power control \cite{Fabregas:CodedMod}. Since a system with short-term power constraints cannot allocate power across
multiple codewords, it is logical that the resulting outage exponent is independent of the quality of CSIT.
Increasing $\dpeak$ subsequently leads to an improvement in the
outage performance. However, when $\dpeak$ exceeds a certain threshold,
there is no extra diversity gain
by increasing $\dpeak$ further (the diversity gain is ``saturated'' due to the limitation on the accuracy of the CSIT).
In other words, a stringent constraint on the peak power
exponent leads to a lot more pronounced detrimental effect in the
case of accurate CSIT (large $\de$) than in the case of very noisy CSIT (small $\de$).

In the limiting case $\de\downarrow 0$, i.e., very noisy CSIT, we have $d(R, d_\text
e, \dpeak)\to m\dsb$,
which is again exactly the outage exponent when there is no CSIT
\cite{Fabregas:CodedMod}. In this case the outage exponent is also independent of $\dpeak$, because the transmitter always uses a constant power in the order of $\SNR^1$.
The case $\de\downarrow 0$ also represents the scenarios in some practical systems in which the CSIT noise variance does not decay with the SNR.
If the CSIT noise variance has such an ``error floor'' in the high-SNR regime, then no extra
diversity gain can be obtained from power control.

On the other
hand, in case $\de\to\infty$, i.e. when the CSIT noise
variance decays exponentially or faster with the SNR, then $d(R,
\de)\to\infty$, $\forall R<M$, as long as the peak exponent constraint is also relaxed to satisfy $\dpeak > 1 + m\dsb\de$. For strictly positive and
finite $\de$, using power control, even with noisy CSIT,
provides an extra diversity gain of $\bigl(m\dsb\bigr)^2\de$ compared to the
no-CSIT case, as long as the peak power constraint is sufficiently relaxed. The presence of the factor $m^2$ also parallels with
the diversity--multiplexing tradeoff result obtained in
\cite{Kim:IT_DMT_TDD} for MIMO channels with Gaussian inputs.

We also learn from the analysis in Appendix~\ref{apdx:main_proof} that at high SNR, when $\dpeak$ is sufficiently large, the
dominant outage event occurs when exactly
$\left\lceil\frac{BR}{M}\right\rceil-1$ of the channel gain
estimates $\hat\gamma_i$'s are much larger than the noise variance
$\sigma_e^2$, and the remaining
$B-\left\lceil\frac{BR}{M}\right\rceil+1$ channel estimates have
the same order of magnitude as $\sigma_\text e^2$. For example,
when the rate is sufficiently small such that $BR\le M$ then a
typical outage event occurs when \emph{all} $B$ channel estimates
are in the order of the CSIT noise variance, leading to the
maximum diversity gain of $mB(1+mB\de)$.
When $\dpeak$ is sufficiently small, however, the system cannot ``invert'' the worst channel realizations and
the peak exponent becomes the limiting factor. For example when $\dpeak<\de$ then the dominant outage event happens even when all
the channel estimates are very accurate (significantly above the CSIT noise level).

\subsection{Improving the Outage Exponent with Rotations}

In \cite{Fabregas:Rotation}, it is shown that a simple precoding technique can be used to improve the outage exponent
over fading channels with discrete inputs and uniform power allocation. In this section, we demonstrate how the idea in \cite{Fabregas:Rotation} can be applied in the current noisy CSIT
setting of interest to further improve the outage exponents. In order to avoid cumbersome notation and to simplify the presentation, we remove the peak exponent constraint (setting $\dpeak=\infty$), focusing only on the effects of the CSIT noise.

\begin{figure}[t]
\begin{center}
\includegraphics[width=0.7\columnwidth]{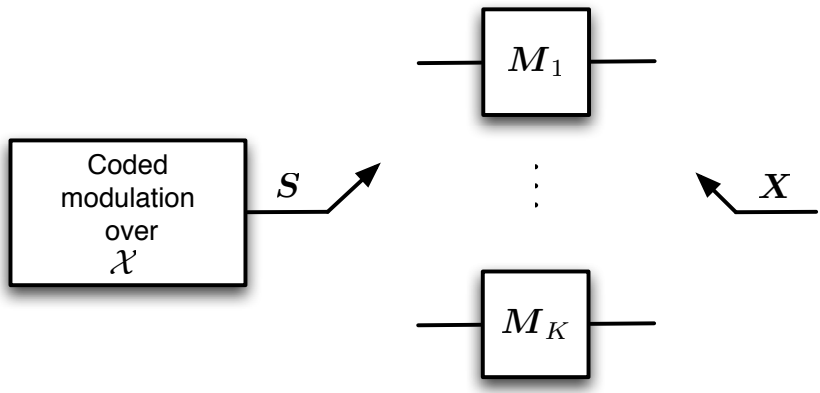}
\caption{Code construction with rotations.}
\label{fig:model_rotations}
\end{center}
\end{figure}

In the following we briefly recall the precoding technique of \cite{Fabregas:Rotation}. First consider reformatting the codewords $\vect x \in \cal M$ as matrices
\beq
\mat X = \begin{pmatrix}
{\vect x}_1\\
\vdots\\
{\vect x}_B
\end{pmatrix}\in\CC^{B\times L}.
\eeq
We now obtain $\mat X$ as
\beq
\mat X = \mat M \mat S
\label{eq:rotation}
\eeq
where
\beq
\mat M = \begin{pmatrix}
{\mat M}_1 &\mat 0&\mat0\\
\mat 0 & \ddots &\mat0\\
\mat 0 & \mat 0 &{\mat M}_K
\end{pmatrix} \in \CC^{B\times B}
\eeq
is a unitary block-diagonal matrix, and the entries of $\mat S\in\CC^{B\times L}$ belong to the signal constellation $\cal X$ with size $2^M$ symbols. The matrices $\mat M_1,\dotsc,\mat M_K \in \CC^{N\times N}$ are the $K$ unitary rotation matrices of dimension $N$ each. Fig. \ref{fig:model_rotations} illustrates the above construction. These rotation matrices are required to have full diversity, i.e.,
\beq
\mat M_k (\vect s - \vect s') \neq \vect 0
\eeq
componentwise, for all $\vect x\neq\vect x' \in \Xc^N$. This implies that if the vector $(\vect s - \vect s')$ has a positive number of nonzero entries, then, its rotated version will have all $N$ entries different from zero. The reader is referred to \cite{Fabregas:Rotation} for more details on the construction and to \cite{Oggier:Foundation2004} for a detailed discussion on the design of full-diversity rotation methods.

According to
\cite{Fabregas:Rotation}, with no CSIT we obtain the following exponent
\begin{equation}
\dout= m \dsbrot
\end{equation}
where
\beq
\dsbrot \define N\left(1 + \left\lfloor \frac{B}{N}\left(1-\frac{R}{M}\right) \right\rfloor\right) = B + N -N\left\lceil\frac{BR}{MN}\right\rceil.
\eeq

With noisy CSIT, completely similarly to the previous section we have the following result.

\begin{theorem}
Consider transmission at rate $R$ over a block-fading channel described by \eqref{eq:model} with Rayleigh fading with mismatched CSIT modeled by \eqref{eq:mismatched_csit} with inputs obtained as the rotation of a coded modulation scheme over $\Xc$ as described by \eqref{eq:rotation}, using full diversity rotations. The transmitter uses power control with an average power constraint \eqref{eq:average_pc}.
Then, the outage exponents are given by
\begin{equation}
d(R, \de) =
m\dsbrot\left(1+m\dsbrot \de\right).
\end{equation}
\label{theorem:di_rot}
\end{theorem}

\begin{IEEEproof}
See Appendix~\ref{apdx:rot_proof}.
\end{IEEEproof}

\begin{figure}[htp]
\begin{center}
\includegraphics[width=1\columnwidth]{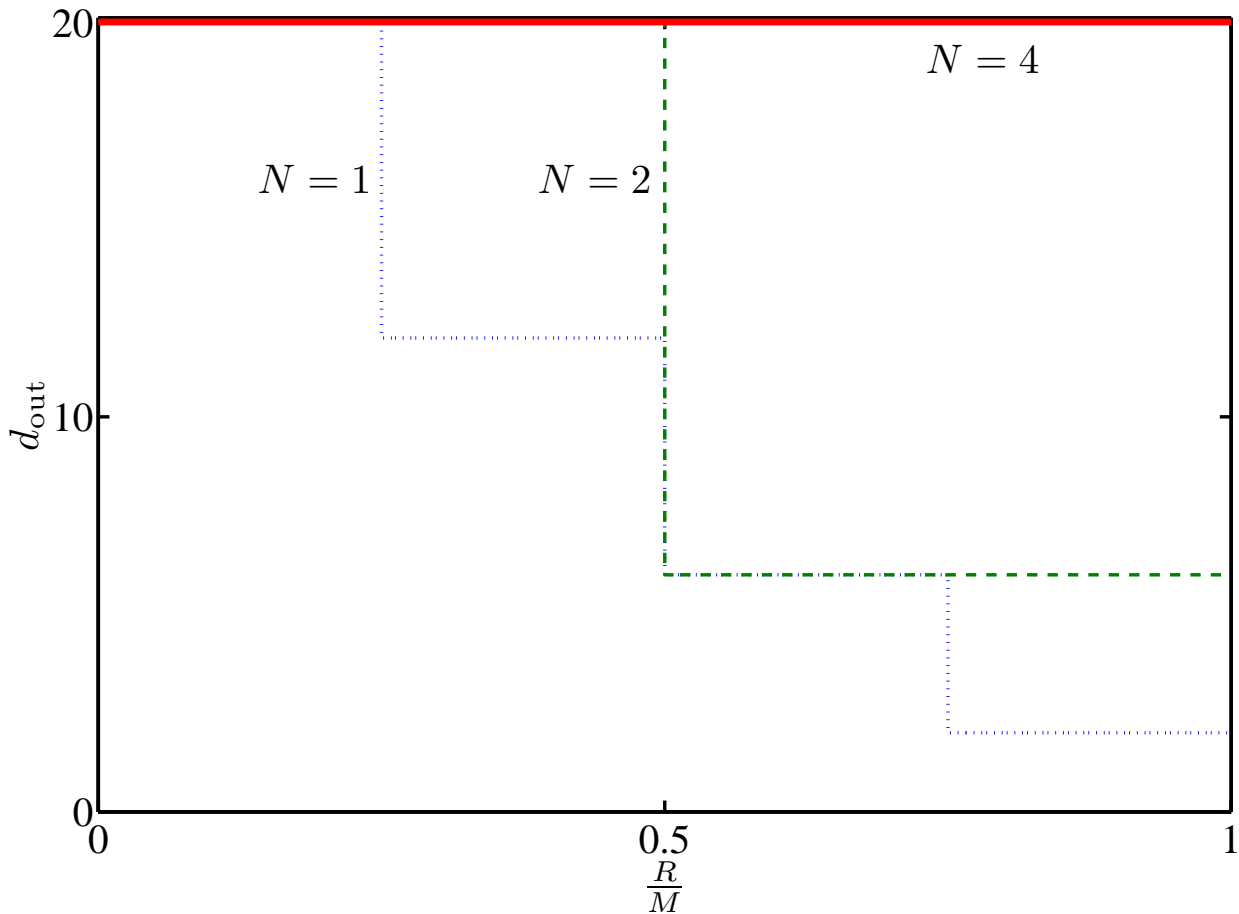}
\caption{Outage exponents for $B=4$, $m=1$, $d_\text e=1$ and full-diversity rotations of size $N=1$ (dotted line), $N=2$ (dashed line) and $N=4$ (solid line).}
\label{fig:exponents_rotations_B4}
\end{center}
\end{figure}

We illustrate in Fig.~\ref{fig:exponents_rotations_B4} the effect of full-diversity rotation matrices on the outage exponent of the coded modulation system with mismatched CSIT. This precoding method clearly leads to a higher diversity gain even at high code rates, at the expense of increasing receiver complexity.

In the special case $N=B$, i.e. when a single matrix that rotates
all $B$ output symbols is used, then $d(R, \de)= mB(1 + mB\de)$.
This is the maximum diversity gain we can achieve in this
scenario, even with codes drawn from a Gaussian ensemble
\cite{Kim:IT_DMT_TDD}. For a large $N$, however, the receiver
complexity will increase exponentially, as this rotation will
require joint decoding, taking the output of blocks of $N$
sub-channels into account. Note also, that, since this strategy
yields the optimal exponent, in terms of exponents, there is
nothing to gain in optimizing the full precoding matrix. Using
power control and a full-dimension full-diversity rotation matrix
is sufficient.

\section{Conclusion}
\label{section:conclusions}
We have studied the asymptotic behavior of the outage probability for code modulation
over block-fading channels under the assumption that the transmitter has access to a noisy version of the instantaneous
channel gains. We showed that power control even with mismatched CSIT is still very beneficial
in improving the outage performance of the system. Our results shed some light into the interplay between different parameters in a coded modulation system, including the constellation size, the code rate, the quality of the CSIT, and the peak power requirement. Determining the outage exponents in a more general multiple-input multiple-output remains an interesting open problem.

\appendices

\section{Proof of Theorem \ref{theorem:di}}\label{apdx:main_proof}

Since we are interested in the high-SNR regime, let us invoke the
standard change of variables as in \cite{Zheng:Tradeoff},
$\bar\alpha_i = -\frac{\log\bar\gamma_i}{\log\SNR}$ and
$\hat\alpha_i = -\frac{\log\hat\gamma_i}{\log\SNR}$. We also
perform the change of variable $\pi(\widehat{\boldsymbol\gamma})
\equiv \pi(\widehat{\boldsymbol\alpha}) \define \frac{\log
P(\widehat{\boldsymbol\gamma})}{\log\SNR}$.

The power constraint
(\ref{eq:average_pc}) asymptotically becomes \cite{Kim:IT_DMT_TDD,Kim:SP_Distortion}
\begin{equation}
\int
\SNR^{\pi(\widehat{\boldsymbol\gamma})}f(\widehat{\boldsymbol\gamma})d\widehat{\boldsymbol\gamma}\dotle
\SNR^1.
\end{equation}
Notice that the $\hat\gamma_i$'s are mutually independent  and
follow Chi-square distribution with $2m$ degrees of freedom. Also,
we have $\expect[\hat\gamma_i] = \expect[\|\vect h_i\|^2] -
\expect[\|\vect e_i\|^2] \doteq \SNR^0$. Changing variables from
$\widehat{\boldsymbol\gamma}$ to $\widehat{\boldsymbol\alpha}$, we
readily obtain
\begin{equation}\int_{\widehat{\boldsymbol\alpha}\in\mathbb R_+^B} \SNR^{\pi(\widehat{\boldsymbol\alpha})}\SNR^{-m\sum_{i=1}^B\hat\alpha_i}d\widehat{\boldsymbol\alpha}\dotle\SNR^1.\end{equation}
Herein we have neglected the terms irrelevant to the SNR exponent,
noticing that for any set containing $\alpha_i<0$, its probability
measure decays exponentially in SNR \cite{Zheng:Tradeoff}.
Applying Varadhan's integral lemma \cite{Dembo:LargeDeviation} we
then have
\begin{equation}\sup_{\widehat{\boldsymbol\alpha}\in\mathbb
R_+^B}\left\{\pi(\widehat{\boldsymbol\alpha})-m\sum_{i=1}^B\hat\alpha_i\right\}\le
1.
\end{equation}
Since outage probability is a non-increasing function of transmit
power, we conclude that with the optimal power allocation,
\begin{equation}
\pi(\widehat{\boldsymbol\alpha}) = \min\left(\dpeak, 1 +
m\sum_{i=1}^B\hat\alpha_i, \right)
\end{equation}
where we need to introduce $\dpeak$ to take into account the peak
constraint (\ref{eq:peak_pc}).

From \cite{Fabregas:CodedMod} it is known that as $\SNR\to\infty$
the mutual information in sub-channel $i$, $I_\mathcal
X\left(P(\widehat{\boldsymbol\gamma})\gamma_i\right)$,
tends to either $M$ or $0$ depending only on the behavior of the
term
\begin{equation}P(\widehat{\boldsymbol\gamma})\gamma_i\doteq\SNR^{\pi(\widehat{\boldsymbol\alpha})}\SNR^{-\de}\SNR^{-\bar\alpha_i}=\SNR^{\min\left(\dpeak, 1+m\sum_{j=1}^B\hat\alpha_j\right)-\de-\bar\alpha_i}.\end{equation}
In particular, if $\bar\alpha_i\le
\pi(\widehat{\boldsymbol\alpha})-\de$ then $I_\mathcal
X\left(P(\widehat{\boldsymbol\gamma})\gamma_i\right)\to
M$ bits per channel use. Otherwise $I_\mathcal
X\left(P(\widehat{\boldsymbol\gamma})\gamma_i\right)\to
0$.

Thus the asymptotic outage set is given by
\begin{equation}\label{eq:out_set}\mathcal O = \left\{\bar{\boldsymbol\alpha},\widehat{\boldsymbol\alpha}:
\sum_{i=1}^B\mathbf 1\left(\bar\alpha_i\le \min\left(\dpeak, 1+m\sum_{j=1}^B\hat\alpha_j\right) - \de\right)<\frac{BR}{M}\right\}
\end{equation}
where $\mathbf 1(\cdot)$ is the
indicator function.
We then have
\begin{equation}\begin{split}P_\text{out}(R)&\doteq\int_\mathcal O f(\bar{\boldsymbol\gamma}|\widehat{\boldsymbol\gamma})f(\widehat{\boldsymbol\gamma})d\bar{\boldsymbol\gamma}d\widehat{\boldsymbol\gamma}\\
&\doteq \int_\mathcal O f(\bar{\boldsymbol\alpha}|\widehat{\boldsymbol\alpha})f(\widehat{\boldsymbol\alpha})d\bar{\boldsymbol\alpha}d\widehat{\boldsymbol\alpha}.
\end{split}\label{eq:gen_pdf}\end{equation}
Notice that
$f(\bar{\boldsymbol\gamma}|\widehat{\boldsymbol\gamma})=\prod_{i=1}^Bf(\bar\gamma_i|\hat\gamma_i)$,
where the conditional p.d.f $f(\bar\gamma_i|\hat\gamma_i)$ is a
non-central chi-square one with $2m$ degrees of freedom. In Appendix~\ref{apdx:asymp_expa_pdf} we asymptotically expand the integral (\ref{eq:gen_pdf}), showing that
the outage exponent is eventually given by \begin{equation}d(R;\de, \dpeak) = \min(d_0, \ldots, d_B)\end{equation}
with $d_n$ being defined such that
\begin{equation}\begin{split}\int_{\mathcal O\cap \mathcal B_n}
\prod_{i=1}^{B-n}\SNR^{-m\hat\alpha_i-m\bar\alpha_i}\prod_{j=B-n+1}^B\SNR^{-m\hat\alpha_j}
d\bar{\boldsymbol \alpha}d\widehat{\boldsymbol
\alpha}\doteq \SNR^{-d_n}\end{split}\end{equation} where
\begin{equation}\begin{split}\mathcal B_n&\define\left\{\bar{\boldsymbol
\alpha},\widehat{\boldsymbol \alpha}:
\{\bar\alpha_1>0,\hat\alpha_1\ge \de\}\cap\cdots\cap
\{\bar\alpha_{B-n}>0,\hat\alpha_{B-n}\ge d_\text
e\}\right.\\
&\quad\left.\cap\{0\le\hat\alpha_{B-n+1}<\de,
\bar\alpha_{B-n+1}=\hat\alpha_{B-n+1}-d_\text
e\}\cap\cdots\cap\{0\le\hat\alpha_{B}<\de,
\bar\alpha_{B}=\hat\alpha_{B}-d_\text
e\}\right\}.\end{split}\end{equation}
Thus applying Varadhan's integral lemma \cite{Dembo:LargeDeviation} gives
\begin{equation}
d_n = \inf_{\bar{\boldsymbol
\alpha},\widehat{\boldsymbol \alpha} \in \mathcal O\cap\mathcal B_n} \left\{m\sum_{i=1}^B\hat\alpha_i + m\sum_{j=1}^{B-n}\bar\alpha_j\right\}.
\label{eq:dn}
\end{equation}

Recall from (\ref{eq:out_set}) that
\begin{equation*}\mathcal O = \left\{\bar{\boldsymbol\alpha},\widehat{\boldsymbol\alpha}:
\sum_{i=1}^B\mathbf 1\left(\bar\alpha_i\le \min\left(\dpeak, 1+m\sum_{j=1}^B\hat\alpha_j\right) - \de\right)<\frac{BR}{M}\right\}.\end{equation*}
Over $\mathcal B_n$, we have that $\bar\alpha_i = \hat\alpha_i -\de$ for all $i \ge B-n+1$, thus
\begin{equation}\begin{split}\mathcal O &= \left\{\bar{\boldsymbol\alpha},\widehat{\boldsymbol\alpha}:
\sum_{i=1}^{B-n}\mathbf 1\left(\bar\alpha_i\le \min\left(\dpeak, 1+m\sum_{j=1}^B\hat\alpha_j\right) - \de\right)\right.\\
&\qquad\qquad\quad\left.+\sum_{i=B-n+1}^{B}\mathbf 1\left(\hat\alpha_i\le \min\left(\dpeak, 1+m\sum_{j=1}^B\hat\alpha_j\right)\right)
<\frac{BR}{M}\right\}.\end{split}
\end{equation}
To compute $d_n$, we consider two mutual exclusively cases.

\textbf{Case 1}: $\dpeak<1 + m\sum_{j=1}^B\hat\alpha_j$. We denote the SNR exponent over the intersection of this region and $\mathcal B_n$ as $d_n^{(1)}$. Then
\begin{equation}\mathcal O = \left\{\bar{\boldsymbol\alpha},\widehat{\boldsymbol\alpha}:
\sum_{i=1}^{B-n}\mathbf 1\left(\bar\alpha_i\le \dpeak - \de\right)
+\sum_{i=B-n+1}^{B}\mathbf 1\left(\hat\alpha_i\le \dpeak\right)
<\frac{BR}{M}\right\}.\end{equation}

\textbf{Case 1.1}: If $\dpeak<\de$ then $\mathbf 1\left(\bar\alpha_i\le \dpeak - \de\right)=0$, $\forall i\in\{1,\ldots,B-n\}$. The outage set reduces to
\begin{equation}\mathcal O = \left\{\widehat{\boldsymbol\alpha}:
\sum_{i=B-n+1}^{B}\mathbf 1\left(\hat\alpha_i\le \dpeak\right)
<\frac{BR}{M}\right\}.\label{eq:O_case11}\end{equation}
Because for $i=1, \ldots, B-n$, the terms $\hat\alpha_i$  and $\bar\alpha_i$ are not present in the outage set,
we have the optimal solution to (\ref{eq:dn}) $\bar\alpha_1^* = \cdots = \bar\alpha_{B-n}^* = 0$ and $\sum_{i=1}^{B-n}\hat\alpha_i^* = \max(\dpeak-1, m(B-n)\de)$, due to the constraint $\dpeak<1 + m\sum_{j=1}^B\hat\alpha_j$.

There are only $n$ terms in the summation in (\ref{eq:O_case11}), thus if $n<\frac{BR}{M}$ then
$$d_n^{(1)} = \max\left(\dpeak-1, m(B-n)\de\right).$$
But since $n<\frac{BR}{M}<B$ and $\dpeak<\de$ we have $m(B-n)\de>m(B-n)\dpeak>\dpeak-1$. Thus
$$d_n^{(1)}=m(B-n)\de$$ if $n<\frac{BR}{M}$.

If $n\ge \frac{BR}{M}$ then without the constraint $\dpeak<1 + m\sum_{j=1}^B\hat\alpha_j$
we readily
obtain the solution to (\ref{eq:dn}): $\hat\alpha_{B-n+1}^* = \cdots = \hat\alpha_{B-\lceil\frac{BR}{M}\rceil+1}^* = \dpeak$ and $\hat\alpha_{B-\lceil\frac{BR}{M}\rceil+2}^* = \cdots = \hat\alpha_{B}^* = 0$.
Taking the constraint $m\sum_{j=1}^B\hat\alpha_j>\dpeak-1$ into account we have
\begin{equation}d_n^{(1)} = \max\left(\dpeak-1, m(B-n)\de+m\dpeak\left(n-\left\lceil\frac{BR}{M}\right\rceil+1\right)\right).\end{equation}
But $\dpeak<\de$ thus we have
\[m(B-n)\de+m\dpeak\left(n-\left\lceil\frac{BR}{M}\right\rceil+1\right) \ge m\dpeak\left(B-\left\lceil\frac{BR}{M}\right\rceil+1\right)\ge m\dpeak>\dpeak-1,\]
and
\begin{equation}d_n^{(1)} = m(B-n)\de+m\dpeak\left(n-\left\lceil\frac{BR}{M}\right\rceil+1\right).\end{equation}
In summary, if $\dpeak<\de$ then
\begin{equation}
d_n^{(1)} = \begin{cases}m(B-n)\de & \text{if\;} \frac{BR}{M}> n,\\
m(B-n)\de+m\dpeak\left(n-\left\lceil\frac{BR}{M}\right\rceil+1\right) & \text{if\;} \frac{BR}{M}\le n.\end{cases}
\label{eq:dn1_qsmall}
\end{equation}

\textbf{Case 1.2}: On the other hand, if $\dpeak\ge \de$, then for $i=B-n+1,\ldots,B$ we have $\mathbf 1(\hat\alpha_i\le \dpeak) = 1$
because in $\mathcal B_n$, $\hat\alpha_i<\de$ for these values of $i$. The outage set reduces to
\begin{equation}\mathcal O = \left\{\bar{\boldsymbol\alpha},\widehat{\boldsymbol\alpha}:
\sum_{i=1}^{B-n}\mathbf 1\left(\bar\alpha_i\le \dpeak - \de\right)
<\frac{BR}{M}-n\right\}.\label{eq:O_case12}
\end{equation}
If $\frac{BR}{M}\le n$ then $d_n^{(1)} = \infty$ because the set of ``bad'' channel realizations is empty \cite{Kim:TradeoffIT}. Intuitively, in this case we have access to
``perfect'' knowledge about $n$ channel gains which we can then use to successfully ``invert'' the channel gain (since $\dpeak$ is sufficiently large and does not pose any restriction). Consequently we can achieve exponential decay in the outage probability for all rates $R\le \frac{Mn}{B}$.

If $\frac{BR}{M}> n$ then, due to the total absence of $\hat\alpha_i$ in (\ref{eq:O_case12}), the optimal solution to (\ref{eq:dn})
satisfies $\sum_{i=1}^B \hat\alpha_i^* = \max(\dpeak-1, m(B-n)\de)$, where we have taken into account the constraint $\dpeak<1 + m\sum_{j=1}^B\hat\alpha_j$.
As for $\bar\alpha_i$'s, from (\ref{eq:O_case12}), we see that at the optimum points, there are exactly $\left\lceil\frac{BR}{M}-n\right\rceil-1$ of the $\bar\alpha_i$'s that are equal to zero, and the remaining $B-n-\left\lceil\frac{BR}{M}-n\right\rceil+1$ variables are all equal (or arbitrarily close to from above, strictly speaking) to $\dpeak-\de$.

Finally we have
\begin{equation}d_n^{(1)} = m(\dpeak-\de)\left(B-n+1-\left\lceil\frac{BR}{M}-n\right\rceil\right) + \max\left(\dpeak-1, m(B-n)\de\right).\end{equation}
In summary, if $\dpeak\ge \de$ then
\begin{equation}d_n^{(1)} =
\begin{cases}m(\dpeak-\de)\left(B-n+1-\left\lceil\frac{BR}{M}-n\right\rceil\right) + \max\left(\dpeak-1, m(B-n)\de\right)
& \text{if\;} \frac{BR}{M}> n\\
\infty & \text{if\;} \frac{BR}{M}\le n.
\end{cases}\label{eq:dn1_qbig}\end{equation}

\textbf{Case 2}: $\dpeak\ge 1 + m\sum_{j=1}^B\hat\alpha_j$. Note that over $\mathcal B_n$ we have $\sum_{j=1}^B\hat\alpha_j \ge (B-n)\de$
thus Case 2 can only happen if $\dpeak\ge 1 + m(B-n)\de$. For $n$ such that $\dpeak< 1 + m(B-n)\de$, we use the convention $d_n^{(2)}=\infty$.
Then, over $\mathcal B_n$
\begin{equation}\begin{split}\mathcal O &= \left\{\bar{\boldsymbol\alpha},\widehat{\boldsymbol\alpha}:
\sum_{i=1}^{B-n}\mathbf 1\left(\bar\alpha_i\le 1+m\sum_{j=1}^B\hat\alpha_j - \de\right)
+\sum_{i=B-n+1}^{B}\mathbf 1\left(\hat\alpha_i\le 1+m\sum_{j=1}^B\hat\alpha_j\right)
<\frac{BR}{M}\right\}\\
&=\left\{\bar{\boldsymbol\alpha},\widehat{\boldsymbol\alpha}:
\sum_{i=1}^{B-n}\mathbf 1\left(\bar\alpha_i\le 1+m\sum_{j=1}^B\hat\alpha_j - \de\right)
<\frac{BR}{M}-n\right\}.
\end{split}\end{equation}
Again if $\frac{BR}{M}\le n$ then the outage event decays exponentially in the SNR.
We readily obtain $\hat\alpha_1^* = \cdots = \hat\alpha_{B-n}^* = \de$
and $\hat\alpha_{B-n+1}^* = \cdots = \hat\alpha_{B}^* = 0$. We also have
$\bar\alpha_i^* = 1+m(B-n)\de - \de$, for exactly $B-n-\left\lceil\frac{BR}{M}-n\right\rceil+1$ of the $\bar\alpha_i$'s, and the other $\bar\alpha_i$'s are zero.
Thus \begin{equation}d_n^{(2)} = \begin{cases} m(B-n)\de + m\left(B-n-\left\lceil\frac{BR}{M}-n\right\rceil+1\right)(1+m(B-n)\de-\de) & \text{if\;}\frac{BR}{M}>n,
\\
\infty & \text{if\;}\frac{BR}{M}\le n.\end{cases}\label{eq:dn2}\end{equation}

We now combine the results in Case 1 and Case 2 to find the outage exponent
\begin{equation}d(R, \de, \dpeak) = \min(d_0, \ldots, d_B) = \min(d_0^{(1)}, d_0^{(2)}, \ldots, d_B^{(1)}, d_B^{(2)}).
\end{equation}

If $\dpeak<\de$ then the $d_n^{(1)}$'s are given by (\ref{eq:dn1_qsmall}). Furthermore, we have $\dpeak<\de<1 + m(B-n)\de$, $\forall n = 0,\ldots, B-1$ thus
$d_n^{(2)} = \infty$ for these values of $n$. For $n=B$, we also have from (\ref{eq:dn2}) that $d_n^{(2)} = \infty$. Thus in this case
\begin{equation}
d(R, \de, \dpeak) = \min(d_0^{(1)}, d_1^{(1)},\ldots, d_B^{(1)}).
\end{equation}

From (\ref{eq:dn1_qsmall}) we have \begin{equation}d_0^{(1)}>d_1^{(1)}>\cdots>d_{\left\lceil\frac{BR}{M}\right\rceil-1}^{(1)} = m\de\left(B - \left\lceil\frac{BR}{M}\right\rceil + 1.\right)\end{equation}
Also from (\ref{eq:dn1_qsmall}) and from the fact that $\dpeak<\de$, we have
\begin{equation}
d_{\left\lceil\frac{BR}{M}\right\rceil}^{(1)} > \cdots > d_B^{(1)} = m\dpeak\left(B-\left\lceil\frac{BR}{M}\right\rceil+1\right).
\end{equation}
Thus we finally have
\begin{equation}\begin{split}d(R, \de, \dpeak) &= \min\left(m\de\left(B - \left\lceil\frac{BR}{M}\right\rceil + 1.\right), m\dpeak\left(B - \left\lceil\frac{BR}{M}\right\rceil + 1.\right)\right)\\
&=m\dpeak\left(B - \left\lceil\frac{BR}{M}\right\rceil + 1.\right)\end{split}\end{equation}
The analysis also reveals that the dominant outage event occurs in the region $\mathcal B_B$, i.e., when \emph{all} the channel estimates
$\hat\gamma_i$'s have a much large order of magnitude than the  CSIT noise variance. More specifically, in the typical outage event, $B-\left\lceil\frac{BR}{M}\right\rceil+1$ of the
channel estimates are in the order of
$\SNR^{-\dpeak}$, canceling out the maximum power that can be allocated to any channel realization.
Thus the limiting factor in this case is the peak
exponent $\dpeak$.

We now consider the case $\dpeak\ge \de$, where the $d_n^{(1)}$'s are given by (\ref{eq:dn1_qbig}). There are
three possibilities.

\textbf{Case A}: If $\dpeak \ge 1 + mB\de$ then $\dpeak\ge 1 + m(B-n)\de$, $\forall n = 0,\ldots, B$. Thus
\begin{equation}d_n^{(1)}=\begin{cases}m(\dpeak-\de)\left(B-n+1-\left\lceil\frac{BR}{M}-n\right\rceil\right) + \dpeak-1
& \text{if\;} \frac{BR}{M}> n\\
\infty & \text{if\;} \frac{BR}{M}\le n.
\end{cases}\end{equation}
But since $\dpeak \ge 1 + m(B-n)\de$\[\begin{split}&m(\dpeak-\de)\left(B-n+1-\left\lceil\frac{BR}{M}-n\right\rceil\right) + \dpeak-1\\
&\quad\ge m(1+m(B-n)\de-\de)\left(B-n+1-\left\lceil\frac{BR}{M}-n\right\rceil\right) + m(B-n)\de.
\end{split}\]
But the right hand side is exactly the value of $d_n^{(2)}$ in (\ref{eq:dn2}) when $\frac{BR}{M}> n$. We conclude that
\[d(R, \de, \dpeak) = \min(d_0^{(2)}, d_1^{(2)}, \ldots, d_B^{(2)}).\]
Furthermore, from (\ref{eq:dn2}) we have that $d_{\left\lceil\frac{BR}{M}\right\rceil-1}^{(2)}<\cdots <d_1^{(2)}<d_0^{(2)}<
 \infty = d_{\left\lceil\frac{BR}{M}\right\rceil}^{(2)} = \cdots = d_B^{(2)}$. Hence
\begin{equation}
\begin{split}d(R, \de, \dpeak) &= d_{\left\lceil\frac{BR}{M}\right\rceil-1}^{(2)}\\
&=m\left(1+m\left(B-\left\lceil\frac{BR}{M}\right\rceil+1\right)\de-\de\right)\left(B-\left\lceil\frac{BR}{M}\right\rceil+2-\left\lceil\frac{BR}{M}-\left\lceil\frac{BR}{M}\right\rceil+1\right\rceil\right) \\
&\quad+ m\left(B-\left\lceil\frac{BR}{M}\right\rceil+1\right)\de\\
&= m\left(B-\left\lceil\frac{BR}{M}\right\rceil+1\right)\left(1+m\left(B-\left\lceil\frac{BR}{M}\right\rceil+1\right)\de\right).
\end{split}
\end{equation}
The dominant outage event occurs when exactly $B-\lceil\frac{BR}{M}\rceil+1$ of the channel gains have the same order of magnitude as the CSIT noise variance. The peak exponent constraint is not the limiting factor in this case.

\textbf{Case B}: $1 + md_\text
e\left(B-\left\lceil\frac{BR}{M}\right\rceil+1\right)< \dpeak<1 +
mB\de$. This implies $\frac{BR}{M}\ge
\left\lceil\frac{BR}{M}\right\rceil-1> B-\frac{\dpeak-1}{m\de}$.
For any integer $n$ such that $n<B-\frac{\dpeak-1}{m\de}$ then
$\dpeak<1+m\de(B-n)$. Thus for these values of $n$, we have
\[d_n^{(1)}=m(\dpeak-\de)\left(B-n+1-\left\lceil\frac{BR}{M}-n\right\rceil\right) + m\de(B-n)\]
and $d_n^{(2)}=\infty$.

As for $n = \left\lceil B-\frac{\dpeak-1}{m\de}\right\rceil, \ldots, \left\lceil\frac{BR}{M}\right\rceil-1$, then $\dpeak\ge 1+m\de(B-n)$.
This is similar to Case A, i.e., we have $d_n^{(1)} \ge d_n^{(2)}$.

Thus in Case B
\[d(R, \de, \dpeak) = \min\left(d_{0}^{(1)}, \ldots, d_{\left\lceil B-\frac{\dpeak-1}{m\de}\right\rceil-1}^{(1)},
d_{\left\lceil B-\frac{\dpeak-1}{m\de}\right\rceil}^{(2)},
\ldots, d_{\left\lceil\frac{BR}{M}\right\rceil-1}^{(2)}\right)\]
due to the fact that $d^{(k)}_{\left\lceil\frac{BR}{M}\right\rceil} =
\cdots = d^{(k)}_B = \infty$ for any $k$. It is readily verifiable that
$d_{0}^{(1)}> \cdots> d_{\left\lceil B-\frac{\dpeak-1}{md_\text
e}\right\rceil-1}^{(1)}> d_{\left\lceil B-\frac{\dpeak-1}{md_\text
e}\right\rceil}^{(2)}> \cdots
> d_{\left\lceil\frac{BR}{M}\right\rceil-1}^{(2)}$ and thus
\begin{equation}\begin{split}d(R, \de, \dpeak) &=
d_{\left\lceil\frac{BR}{M}\right\rceil-1}^{(2)}\\
&=m\left(B-\left\lceil\frac{BR}{M}\right\rceil+1\right)\left(1+m\left(B-\left\lceil\frac{BR}{M}\right\rceil+1\right)d_\text
e\right).
\end{split}\end{equation}
The dominant outage event also happens when $B-\lceil\frac{BR}{M}\rceil+1$ of the channel gains have the same order of magnitude as the CSIT noise variance.

\textbf{Case C}: $ \dpeak \le 1 + md_\text
e\left(B-\left\lceil\frac{BR}{M}\right\rceil +1\right)$. This
implies $\left\lceil\frac{BR}{M}\right\rceil-1\le
B-\frac{\dpeak-1}{m\de}$. Thus for any integer $n$ such that
$n<\frac{BR}{M}$ then $n<B-\frac{\dpeak-1}{m\de}$ leading to $\dpeak<1
+ m\de(B-n)$. Hence from (\ref{eq:dn1_qbig}) we  have
\begin{equation}d_n^{(1)}=\begin{cases}m(\dpeak-\de)\left(B-n+1-\left\lceil\frac{BR}{M}-n\right\rceil\right) + m\de(B-n)
& \text{if\;} \frac{BR}{M}> n\\
\infty & \text{if\;} \frac{BR}{M}\le n.
\end{cases}\end{equation}
Since $n<\frac{BR}{M}$ leads to $\dpeak<1 + m\de(B-n)$, we also have $d_n^{(2)}=\infty$, $\forall n$.
Thus
\begin{equation}
\begin{split}
d(R, \de, \dpeak) &= \min(d_0^{(1)}, \ldots, d_B^{(1)})\\
&= d_{\left\lceil\frac{BR}{M}\right\rceil - 1}^{(1)}\\
&=m(\dpeak-\de)\left(B-\left\lceil\frac{BR}{M}\right\rceil +2-\left\lceil\frac{BR}{M}-\left\lceil\frac{BR}{M}\right\rceil + 1\right\rceil\right)\\
&\quad+ m\de\left(B-\left\lceil\frac{BR}{M}\right\rceil + 1\right)\\
&= m\dpeak\left(B-\left\lceil\frac{BR}{M}\right\rceil + 1\right).
\end{split}
\end{equation}
Again the dominant outage event occurs when exactly $B-\lceil\frac{BR}{M}\rceil+1$ of the channel estimates have the same the order
of magnitude as the CSIT noise variance. Unlike in Case A and Case B, in this case the peak exponent $\dpeak$ is too small and becomes the factor preventing the system from
achieving its full potential. \hfill\IEEEQED

\section{Asymptotic Expansion of (\ref{eq:gen_pdf})}\label{apdx:asymp_expa_pdf}

In this appendix we review the asymptotic expansion of the joint p.d.f. in (\ref{eq:gen_pdf}), a result derived in \cite{Kim:IT_DMT_TDD}
for the case of a single fading block. In particular
we would like to study the high-$\SNR$ behavior of
\begin{equation}
P_\text{out}(R)\doteq\int_\mathcal O \prod_{i=1}^Bf(\bar\gamma_i|\hat\gamma_i)d\bar\gamma_i d\hat\gamma_i.
\end{equation}
where $f(\bar\gamma_i|\hat\gamma_i)$ is a
non-central chi-square p.d.f with $2m$ degrees of freedom and non-centrality parameter
$\frac{2\hat\gamma_i}{\sigma^2_{\text e}}\doteq \SNR^{-\hat\alpha_i+d_{\text e}}$.
Changing variables to
$\widehat{\boldsymbol\alpha}$ and $\bar{\boldsymbol\alpha}$ gives
\begin{equation}\begin{split}
P_\text{out}(R)&\doteq\int_\mathcal O \prod_{i=1}^B
e^{-\SNR^{-\bar\alpha_i}}e^{-\SNR^{-(\hat\alpha_i-d_\text
e)}}e^{-\SNR^{-\hat\alpha_i}}\\
&\quad\times\SNR^{\frac{m-1}{2}(\hat\alpha_i-\bar\alpha_i-d_\text
e) - \bar\alpha_i-m\hat\alpha_i} I_{m-1}\left(\SNR^{\frac{d_\text
e-\bar\alpha_i-\hat\alpha_i}{2}}\right)
d\bar\alpha_id\hat\alpha_i.
\end{split}\label{eq:Pout_bigform}\end{equation}
For each $i\in\{1,\ldots,B\}$, let us define the set
\begin{equation}\mathcal A_i^{(0)}\define
\left\{\hat\alpha_i,\bar\alpha_i: d_\text
e-\hat\alpha_i-\bar\alpha_i>0 \right\}\end{equation} and its
complement
\begin{equation}\mathcal A_i^{(1)}\define
\left\{\hat\alpha_i,\bar\alpha_i: d_\text
e-\hat\alpha_i-\bar\alpha_i\le 0 \right\}.\end{equation}


Firstly, consider the region $\mathcal A_i^{(0)}$, i.e., $d_\text
e-\hat\alpha_i-\bar\alpha_i>0$, for some $i$. Then $\SNR^{d_\text
e-\hat\alpha_i-\bar\alpha_i}\to\infty$ as $\SNR\to\infty$. But for
real $x>0$ we have \cite[Sec. 9.7]{Abramowitz_Stegun}
\begin{equation}I_{m-1}(x) = \frac{e^x}{\sqrt{2\pi x}}\left(1 + O(1/x)\right)\end{equation} thus
$I_{m-1}\left(\SNR^{\frac{d_\text
e-\hat\alpha_i-\bar\alpha_i}{2}}\right)\doteq \SNR^{-\frac{d_\text
e-\hat\alpha_i-\bar\alpha_i}{4}}e^{\SNR^{\frac{d_\text
e-\hat\alpha_i-\bar\alpha_i}{2}}}$. Grouping the exponent terms
inside the integral (\ref{eq:Pout_bigform}) gives
\[\exp\left(-\SNR^{-\bar\alpha_i}-\SNR^{-(\hat\alpha_i-d_\text
e)} + \SNR^{\frac{d_\text
e-\hat\alpha_i-\bar\alpha_i}{2}}\right)\exp\left(-\SNR^{-\hat\alpha_i}\right).\]
Note that
\begin{equation}\max(-\bar\alpha_i, -(\hat\alpha_i-\de)) \ge
\frac{\de-\hat\alpha_i-\bar\alpha_i}{2}\end{equation} for
any $\hat\alpha_i, \bar\alpha_i$ with the equality occurring iff
$\bar\alpha_i  = \hat\alpha_i-\de$. Therefore if
$\bar\alpha_i  \neq \hat\alpha_i-\de$ then
\[-\SNR^{-\bar\alpha_i}-\SNR^{-(\hat\alpha_i-d_\text
e)} + \SNR^{\frac{\de-\hat\alpha_i-\bar\alpha_i}{2}} \doteq
-\SNR^{\max\left(-\bar\alpha_i,-(\hat\alpha_i-\de)\right)}\]
But we are considering $\mathcal A_i^{(0)}$ where $d_\text
e-\hat\alpha_i-\bar\alpha_i>0$, so $\max(\de-\hat\alpha_i,
-\bar\alpha_i)>0$. Thus if $\bar\alpha_i  \neq \hat\alpha_i-d_\text
e$ then the outage probability decays exponentially in $\SNR$.

If $\bar\alpha_i  = \hat\alpha_i-\de$ then the condition
$\de-\hat\alpha_i-\bar\alpha_i>0$ leads to
$\hat\alpha_i<\de$. We also have $\SNR^{\bar\alpha_i} =
\SNR^{\hat\alpha_i}\SNR^{-\de}$ or $\bar\gamma_i =
\hat\gamma_i\sigma^2_\text e$. Thus we can  write
\begin{equation}\begin{split}\int_{\mathcal O\cap\mathcal A_i^{(0)}}
g_i f(\bar\gamma_i|\hat\gamma_i)f(\hat\gamma_i)d\bar\gamma_i
d\hat\gamma_i &\doteq \int_{\mathcal O\cap\{\hat\alpha_i<d_\text
e, \bar\alpha_i=\hat\alpha_i-\de\}}
g_i f(\hat\gamma_i)d\hat\gamma_i
\\
&\doteq \int_{\mathcal O\cap\{\hat\alpha_i<\de,
\bar\alpha_i=\hat\alpha_i-\de\}}
g_i e^{-\SNR^{-\hat\alpha_i}}\SNR^{-m\hat\alpha_i}d\hat\alpha_i
\\
&\doteq \int_{\mathcal O\cap\{0\le\hat\alpha_i<\de,
\bar\alpha_i=\hat\alpha_i-\de\}}
g_i\SNR^{-m\hat\alpha_i}d\hat\alpha_i.\end{split}\label{eq:reg_A0}\end{equation}
Herein we have denoted
\begin{equation}
g_i = \prod_{j=1, j\neq i}^B f(\bar\alpha_j|\hat\alpha_j)d\bar\alpha_jd\hat\alpha_j.
\end{equation}


Secondly, consider the region $\mathcal A_i^{(1)}$ where $d_\text
e-\bar\alpha_i-\hat\alpha_i\le 0$ and thus the asymptotic form of
the modified Bessel function of the first kind $I_{m-1}(x)$ with
$x\downarrow 0$ gives
\begin{equation}I_{m-1}\left(\SNR^{\frac{d_\text
e-\bar\alpha_i-\hat\alpha_i}{2}}\right)\doteq\SNR^{(m-1)\frac{d_\text
e-\bar\alpha_i-\hat\alpha_i}{2}}.\end{equation} We can then
constrain $\bar\alpha_i\ge 0$ and $\hat\alpha_i\ge \de$,
because otherwise the outage probability decays exponentially.
Thus (cf. (\ref{eq:Pout_bigform}))
\begin{equation}
\int_{\mathcal O\cap\mathcal
A_i^{(1)}}g_i f(\bar\gamma_i|\hat\gamma_i)f(\hat\gamma_i)d\hat\gamma_id\bar\gamma_i\doteq\int_{\mathcal
O\cap\{\bar\alpha_i\ge 0,\hat\alpha_i\ge d_\text
e\}}g_i\SNR^{-m\bar\alpha_i -
m\hat\alpha_i}d\bar\alpha_id\hat\alpha_i.
\label{eq:reg_A1}\end{equation}
Recall that $g_i$ collects all the terms that are independent of  $\alpha_i$ and $\hat\alpha_i$.

Thus in the asymptotic expansion of the outage probability, we need to consider $2^B$ regions $\cap_{i=1}^B \mathcal A_i^{(c_i)}\cap \mathcal O$ where $c_i = 0, 1$.
The slowest decaying terms among these $2^B$  regions will determine the outage exponent. However, due to complete symmetry, we can assume without loss of
generality that $\hat\alpha_1 \ge \cdots \ge \hat\alpha_B$. Then the number of regions need considering reduces to $B+1$. In particular for each $n\in\{0,\ldots,
B\}$  we need to
find the exponent $d_n$ where
\begin{equation}\int_{\mathcal O\cap\{\hat\alpha_0 \ge \cdots \ge
\hat\alpha_{B-n} \ge \de
> \hat\alpha_{B-n+1} \ge \cdots \ge \hat\alpha_{B+1}\}}f(\bar{\boldsymbol \alpha}|\widehat{\boldsymbol \alpha})f(\widehat{\boldsymbol \alpha})
d\bar{\boldsymbol \alpha}d\widehat{\boldsymbol
\alpha}\doteq\SNR^{-d_n}
\end{equation}
with the convention $\hat\alpha_0 = \infty$ and
$\hat\alpha_{B+1}=-\infty$. Then $d(R, \de, \dpeak) =
\min(d_0,\ldots, d_B)$.

From (\ref{eq:reg_A0}) and (\ref{eq:reg_A1}), we have
\begin{equation}\begin{split}&\quad\int_{\mathcal O\cap\{\hat\alpha_0 \ge \cdots \ge
\alpha_{B-n} \ge \de
> \alpha_{B-n+1} \ge \cdots \ge \alpha_{B+1}\}}f(\bar{\boldsymbol \alpha}|\widehat{\boldsymbol \alpha})f(\widehat{\boldsymbol \alpha})
d\bar{\boldsymbol \alpha}d\widehat{\boldsymbol \alpha}\\
&\doteq \int_{\mathcal O\cap \mathcal B_n}
\prod_{i=1}^{B-n}\SNR^{-m\hat\alpha_i-m\bar\alpha_i}\prod_{j=B-n+1}^B\SNR^{-m\hat\alpha_j}
d\bar{\boldsymbol \alpha}d\widehat{\boldsymbol
\alpha}\end{split}\end{equation} where
\begin{equation}\begin{split}\mathcal B_n&\define\left\{\bar{\boldsymbol
\alpha},\widehat{\boldsymbol \alpha}:
\{\bar\alpha_1>0,\hat\alpha_1\ge \de\}\cap\cdots\cap
\{\bar\alpha_{B-n}>0,\hat\alpha_{B-n}\ge d_\text
e\}\right.\\
&\quad\left.\cap\{0\le\hat\alpha_{B-n+1}<\de,
\bar\alpha_{B-n+1}=\hat\alpha_{B-n+1}-d_\text
e\}\cap\cdots\cap\{0\le\hat\alpha_{B}<\de,
\bar\alpha_{B}=\hat\alpha_{B}-d_\text
e\}\right\}.\end{split}\end{equation}\hfill\IEEEQED

\section{Proof of Theorem \ref{theorem:di_rot}}\label{apdx:rot_proof}

Similarly to the previous proof we have,

\begin{equation}d(R, \de) = \min(d_0, \ldots, d_B)\end{equation}
where
\begin{equation}d_n = \begin{cases}\infty & \text{if\;} \left\lceil\frac{n}{N}\right\rceil\ge\frac{BR}{MN},\\
mN\left(\frac{B}{N}-\lceil\frac{n}{N}\rceil-K_n\right)(1+m(B-n)\de-\de) + m(B-n)\de &
\text{if\;}\left\lceil\frac{n}{N}\right\rceil<\frac{BR}{MN}.\end{cases}\end{equation} Herein
\begin{equation}K_n = \left\lceil\frac{BR}{MN}-\left\lceil\frac{n}{N}\right\rceil\right\rceil - 1.\end{equation}
In this case $d_n$ is the dominant outage exponent conditioned on the event
that there are exactly $n$ channel gain estimates $\hat\gamma_i$
having a larger order of magnitude than the CSIT noise variance
$\sigma_\text e^2$. Note that by definition $d_i  = \infty$,
$\forall i : \left\lceil\frac{i}{N}\right\rceil\ge \frac{BR}{MN}$. Intuitively, when at least
$i$ channel gains are known
(asymptotically) noiselessly at the transmitter, then using power
control we can always transmit
$MN\times\left\lceil\frac{i}{N}\right\rceil \ge BR$
bits with exponentially decaying error probability. This is because at worst, these known channel gains belong to the least
number of rotation groups, which is $\left\lceil\frac{i}{N}\right\rceil$.

Then from the definition of $d_n$ we have
$d_{N\left(\left\lceil\frac{BR}{MN}\right\rceil-1\right)}\le
d_{N\left(\left\lceil\frac{BR}{MN}\right\rceil-1\right)-1}\le \cdots \le d_0$ and
$d_{N\left(\left\lceil\frac{BR}{MN}\right\rceil-1\right)+1} = \cdots = d_B =
\infty$. Thus
\begin{equation}\begin{split}d(R,\de)&=\min(d_0,\ldots, d_B)\\
&= d_{N\left(\left\lceil\frac{BR}{MN}\right\rceil-1\right)}\\
&=m\left(B-N\left(\left\lceil\frac{BR}{MN}\right\rceil-1\right)\right)d_\text
e\\
&\quad+mN\left(\frac{B}{N}-\left\lceil\frac{BR}{MN}\right\rceil+1-K_{N\left(\left\lceil\frac{BR}{MN}\right\rceil-1\right)}\right)
\left(1+m\left(B-N\left(\left\lceil\frac{BR}{MN}\right\rceil-1\right)\right)d_\text
e-\de\right)\\
&=m\left(B+N-N\left\lceil\frac{BR}{MN}\right\rceil\right)d_\text
e\\
&\quad+mN\left(\frac{B}{N}-\left\lceil\frac{BR}{MN}\right\rceil+2-\left\lceil\frac{BR}{MN}-\left\lceil\frac{BR}{MN}\right\rceil+1\right\rceil\right)
\left(1+m\left(B+N-N\left\lceil\frac{BR}{MN}\right\rceil\right)d_\text
e -\de\right)\\
&=m\left(B+N-N\left\lceil\frac{BR}{MN}\right\rceil\right)d_\text
e\\
&\quad+mN\left(\frac{B}{N}-\left\lceil\frac{BR}{MN}\right\rceil+1\right)
\left(1+m\left(B+N-N\left\lceil\frac{BR}{MN}\right\rceil\right)d_\text
e - \de\right)\\
&=m\left(B+N-N\left\lceil\frac{BR}{MN}\right\rceil\right)\left(1 +
m\left(B+N-N\left\lceil\frac{BR}{MN}\right\rceil\right)\de
\right).\end{split}\end{equation}\hfill\IEEEQED


\begin{thebibliography}{10}
\providecommand{\url}[1]{#1}
\csname url@samestyle\endcsname
\providecommand{\newblock}{\relax}
\providecommand{\bibinfo}[2]{#2}
\providecommand{\BIBentrySTDinterwordspacing}{\spaceskip=0pt\relax}
\providecommand{\BIBentryALTinterwordstretchfactor}{4}
\providecommand{\BIBentryALTinterwordspacing}{\spaceskip=\fontdimen2\font plus
\BIBentryALTinterwordstretchfactor\fontdimen3\font minus
  \fontdimen4\font\relax}
\providecommand{\BIBforeignlanguage}[2]{{%
\expandafter\ifx\csname l@#1\endcsname\relax
\typeout{** WARNING: IEEEtran.bst: No hyphenation pattern has been}%
\typeout{** loaded for the language `#1'. Using the pattern for}%
\typeout{** the default language instead.}%
\else
\language=\csname l@#1\endcsname
\fi
#2}}
\providecommand{\BIBdecl}{\relax}
\BIBdecl

\bibitem{Caire:PowerControl}
G.~Caire, G.~Taricco, and E.~Biglieri, ``Optimum power control over fading
  channels,'' \emph{{IEEE} Trans. Inf. Theory}, vol.~45, pp. 1468--1489, Jul.
  1999.

\bibitem{Love:MIMOTutorials}
D.~J. Love, R.~W. {Heath Jr.}, W.~Santipach, and M.~L. Honig, ``What is the
  value of limited feedback for {MIMO} channels?'' \emph{{IEEE} Commun. Mag.},
  vol.~42, pp. 54--59, Oct. 2004.

\bibitem{Vu:PrecodingTutorials}
M.~Vu and A.~Paulraj, ``{MIMO} wireless linear precoding,'' \emph{{IEEE} Signal
  Process. Mag.}, vol.~24, pp. 86--105, Sep. 2007.

\bibitem{Lim:TradeoffWC}
A.~Lim and V.~K.~N. Lau, ``On the fundamental tradeoff of spatial diversity and
  spatial multiplexing of {MISO}/{SIMO} links with imperfect {CSIT},''
  \emph{{IEEE} Trans. Wireless Commun.}, vol.~7, pp. 110--117, Jan. 2008.

\bibitem{Visotsky:Precoding}
E.~Visotsky and U.~Madhow, ``Space-time transmit precoding with imperfect
  feedback,'' \emph{{IEEE} Trans. Inf. Theory}, vol.~47, pp. 2632--2639, Sep.
  2001.

\bibitem{Jongren:BF_OSTBC}
G.~J{\"o}ngren, M.~Skoglund, and B.~Ottersten, ``Combining beamforming and
  orthogonal space-time block coding,'' \emph{{IEEE} Trans. Inf. Theory},
  vol.~48, pp. 611--627, Mar. 2002.

\bibitem{Zhou:EigenBF}
S.~Zhou and G.~B. Giannakis, ``Optimal transmitter eigen-beamforming and
  space-time block coding based on channel mean feedback,'' \emph{{IEEE} Trans.
  Signal Process.}, vol.~50, no.~10, pp. 2599--2613, Oct. 2002.

\bibitem{Kim:IT_DMT_TDD}
T.~T. Kim and G.~Caire, ``Diversity gains of power control in {MIMO} channels
  with noisy {CSIT},'' \emph{{IEEE} Trans. Inf. Theory}, pp. 1618--1626, Apr.
  2009.

\bibitem{Kamath:Globecom01}
K.~M. Kamath and D.~L. Goeckel, ``Adaptive modulation schemes for minimum
  outage probability in wireless systems,'' in \emph{Proc. {IEEE} Globecom},
  San Antonio, TX, Nov. 2001, pp. 1267--1271.

\bibitem{Sharma:PowerAllocation}
V.~Sharma, K.~Premkumar, and R.~N. Swamy, ``Exponential diversity achieving
  spatio–temporal power allocation scheme for fading channels,'' \emph{{IEEE}
  Trans. Inf. Theory}, pp. 188--208, Jan. 2008.

\bibitem{Steger:Training}
C.~Steger and A.~Sabharwal, ``Single-input two-way {SIMO} channel:
  diversity-multiplexing tradeoff with two-way training,'' \emph{{IEEE} Trans.
  Wireless Commun.}, pp. 4877--4885, Dec. 2008.

\bibitem{Aggrawal:Asilomar08}
V.~Aggarwal, G.~G. Krishna, S.~Bhashyam, and A.~Sabharwal, ``Two models for
  noisy feedback in {MIMO} channels,'' in \emph{Proc. Asilomar Conf. Signals,
  Systems, Computers}, Pacific Grove, CA, Oct. 2008.

\bibitem{Kim:TradeoffIT}
T.~T. Kim and M.~Skoglund, ``Diversity-multiplexing tradeoff in {MIMO} channels
  with partial {CSIT},'' \emph{{IEEE} Trans. Inf. Theory}, vol.~53, pp.
  2743--2759, Aug. 2007.

\bibitem{Zheng:Tradeoff}
L.~Zheng and D.~N.~C. Tse, ``Diversity and multiplexing: {A} fundamental
  tradeoff in multiple-antenna channels,'' \emph{{IEEE} Trans. Inf. Theory},
  vol.~49, pp. 1073--1096, May 2003.

\bibitem{Ozarow:Cellular}
L.~H. Ozarow, S.~{Shamai (Shitz)}, and A.~D. Wyner, ``Information theoretic
  considerations for cellular mobile radio,'' \emph{{IEEE} Trans. Veh.
  Technol.}, vol.~43, pp. 359--378, May 1994.

\bibitem{Biglieri:Fading}
E.~Biglieri, J.~Proakis, and S.~{Shamai (Shitz)}, ``Fading channels:
  {I}nformation-theoretic and communications aspects,'' \emph{{IEEE} Trans.
  Inf. Theory}, vol.~44, pp. 2619--2692, Oct. 1998.

\bibitem{Fabregas:CodedMod}
A.~{Guill{\'e}n i F{\`a}bregas} and G.~Caire, ``Coded modulation in the
  block-fading channel: {C}oding theorems and code construction,'' \emph{{IEEE}
  Trans. Inf. Theory}, vol.~52, pp. 91--114, Jan. 2006.

\bibitem{Nguyen:TightLB}
K.~D. Nguyen, A.~{Guill\'en i F\`abregas}, and L.~K. Rasmussen, ``A tight lower
  bound to the outage probability of discrete-input block-fading channels,''
  \emph{{IEEE} Trans. Inf. Theory}, vol.~53, pp. 4314--4322, Nov. 2007.

\bibitem{Knopp:CodedModulation}
R.~Knopp and P.~A. Humblet, ``On coding for block fading channels,''
  \emph{{IEEE} Trans. Inf. Theory}, vol.~46, pp. 189--205, Jan. 2000.

\bibitem{Malkamaki:CodedDiversity}
E.~Malkam{\"a}ki and H.~Leib, ``Coded diversity on block-fading channels,''
  \emph{{IEEE} Trans. Inf. Theory}, vol.~45, pp. 771--781, Mar. 1999.

\bibitem{Nguyen:ISIT2008}
K.~D. Nguyen, A.~{Guill{\'e}n i F{\`a}bregas}, and L.~K. Rasmussen,
  ``Asymptotic outage performance of power allocation in block-fading
  channels,'' in \emph{Proc. {IEEE} Int. Symp. Information Theory}, Toronto,
  Canada, Jul. 2008, pp. 275--279.

\bibitem{Fabregas:Rotation}
A.~{Guill{\'e}n i F{\`a}bregas} and G.~Caire, ``Multidimensional coded
  modulation in block-fading channels,'' \emph{{IEEE} Trans. Inf. Theory},
  vol.~54, pp. 2367--2372, May 2008.

\bibitem{Oggier:Foundation2004}
F.~Oggier and E.~Viterbo, ``Algebraic number theory and code design for
  {Rayleigh} fading channels,'' \emph{Foundations and Trends in Communications
  and Information Theory}, vol.~1, pp. 333--415, 2004.

\bibitem{Kim:SP_Distortion}
T.~T. Kim, M.~Skoglund, and G.~Caire, ``On source transmission over {MIMO}
  channels with limited feedback,'' \emph{{IEEE} Trans. Signal Process.},
  vol.~57, pp. 324--341, Jan. 2009.

\bibitem{Dembo:LargeDeviation}
A.~Dembo and O.~Zeitouni, \emph{Large Deviations Techniques and
  Applications}.\hskip 1em plus 0.5em minus 0.4em\relax New York: Springer,
  1998.

\bibitem{Abramowitz_Stegun}
M.~Abramowitz and I.~A. Stegun, \emph{Handbook of Mathematical Functions with
  Formulas, Graphs, and Mathematical Tables}.\hskip 1em plus 0.5em minus
  0.4em\relax New York: Dover, 1964.

\end{thebibliography}
\end{document}